\documentclass[preprint, 5p, times, utf8]{elsarticle}
\usepackage{amsmath}
\usepackage{graphicx}			
\usepackage{graphics}
\usepackage{natbib}
\usepackage{textcomp}
\usepackage{gensymb}
\usepackage[hidelinks]{hyperref}
\usepackage{xcolor}
\usepackage{siunitx}
\usepackage{mathrsfs}

\newcommand{\figref}[1]{Fig. \ref{#1}}

\title{Effects of neuronal variability on phase synchronization of neural networks}
\author[ufpr]{K. L. Rossi \corref{cor1}}
\author[ufpr]{R. C. Budzinski}
\author[ufpr]{J. A. P. Silveira}
\author[ufpr]{B. R. R. Boaretto}
\author[ufpr]{T. L. Prado}
\author[ufpr]{S. R. Lopes}\ead{lopes@fisica.ufpr.br}
\author[icbme]{U. Feudel}
\address[ufpr]{Department of Physics, Universidade Federal do Paran\'a, 81531-980 Curitiba, Brazil}
\address[icbme]{Theoretical Physics/Complex Systems, ICBM, Carl von Ossietzky University Oldenburg, 26111 Oldenburg, Germany}
\cortext[cor1]{Corresponding author}

\journal{Neural Networks}

\begin{document}

\begin{frontmatter}
  
\begin{abstract}
An important idea in neural information processing is the communication-through-coherence hypothesis, according to which communication between two brain regions is effective only if they are phase-locked. 
Also of importance is neuronal variability, a phenomenon in which a single neuron's inter-firing times may be highly variable.
In this work, we aim to connect these two ideas by studying the effects of that variability on the capability of neurons to reach phase synchronization.
We simulate a network of modified-Hodgkin-Huxley-bursting neurons possessing a small-world topology. 
First, variability is shown to be correlated with the average degree of phase synchronization of the network. 
Next, restricting to spatial variability - which measures the deviation of firing times between all neurons in the network - we show that it is positively correlated to a behavior we call promiscuity, which is the tendency of neurons to to have their relative phases change with time.
This relation is observed in all cases we tested, regardless of the degree of synchronization or the strength of the inter-neuronal coupling: high variability implies high promiscuity (low duration of phase-locking), even if the network as a whole is synchronized and the coupling is strong.
We argue that spatial variability actually generates promiscuity. Therefore, we conclude that variability has a strong influence on both the degree and the manner in which neurons phase synchronize, which is another reason for its relevance in neural communication. 
\end{abstract}


\end{frontmatter}

\section{Introduction}
\label{sec:intro}

The brain manages to robustly and efficiently process huge amounts of information \cite{Deneve2017, DeMarse2001, Neumann:1958:CB:578873}.
Of the various different ways in which this information can be encoded, one then expects that they exhibit at least some degree of robustness. 
Two possible coding strategies are the rate and temporal codes, in which information is encoded in the rate and in the timing of spikes, respectively \cite{Dayan2005}. Considering that neurons can have variations in their inter-firing intervals \cite{Shadlen3870, NOGUEIRA2018170, NAWROT2008374, Vogels1989, Churchland2010, TOLHURST1983775, KARA2000}, it is important to consider the interactions between this variability and the supposed robustness of the two codes. The neuronal variability may represent unwanted noise insofar as it causes fluctuations in the firing rate \cite{Stein2005}. In this case, for a constant input, one would expect a constant firing rate coding it, which may in reality not happen because of the variability. On the other hand, variability may represent extra information since it affects the timing of spikes \cite{Stein2005}, in which case the differences in the timings, generated by the variability, can carry information.

Besides these effects on the codes, variability may also have an indirect effect on information processing because it can influence the phase synchronization of networks. This phenomenon of phase synchronization happens when neurons start their firing at or sufficiently close to the same time. It is related to various memory and conscious processes \cite{Fell2011, Gaillard_consciousProcessing, Dehaene2014} and is also important in information processing, where it serves to increase the saliency of neural responses, thus facilitating their linking (binding)  \cite{Fell2011, Singer1999, CNC_FRIES}.

Another important example relating oscillatory behavior to information coding is given by Fries \cite{CNC_FRIES}. It relies on the observation that neuronal groups have a tendency to oscillate and that these oscillations affect the likelihood of spike output and sensitivity to synaptic input. The author proposes, that these oscillations create windows of effective communication, when regions have their output and input sensitivity coinciding \cite{CNC_FRIES}.  
Further, a phase can be defined for an oscillating time series to represent where in the oscillation cycle the series is. In this way, coherently oscillating regions can be viewed as a phase-locked (their relative phase is kept constant). With this terminology, we can rephrase the communication-through-coherence hypothesis stated above to say that communication is effective only between phase-locked regions.

The effect that neural variability has on phase synchronization and phase-locking is still unclear.
In this work, we investigate this relation by simulating a network of temperature-dependent  Hodgkin-Huxley-type neurons, proposed by Braun \textit{et al.} \cite{braun1998computer}, and henceforth called HB neurons, coupled in a small-world and random topologies with excitatory chemical synapses.

A network topology, also sometimes called connection scheme, refers to the structure in which neurons are connected. A small-world topology is characterized by having a short average path length (distance) between neurons, yet with high local clustering \cite{watts_1998}. From a neuronal information processing point of view, this is very appealing since it supports both segregated (local) and distributed (global) behaviors \cite{SW_brain_networks}. Small-world networks also have been shown to have high efficienciy, due to having a high complexity even with only few nonlocal connections \cite{SW_efficiency}, and to optimize wiring and energy costs \cite{SW_brain_networks, SW_optimal_wiring}. Supporting this theoretical attractiveness, these topologies have been observed in a wide range of cases, such as in the nervous system of \textit{C. elegans} \cite{watts_1998}, the macaque monkey, the cat \cite{Hilgetag2000}, and in the vertebrate brainstem reticular formation \cite{Humphries2006}.

Neurons in the HB model present bursts, which are rapid sequences of spikes followed by a long quiescent period.
Bursting neurons are thought to be important in information processing \cite{Swadlow2001}: compared to single spikes, bursts are more reliably transmitted to postsynaptic neurons \cite{LISMAN199738} and appear to transmit at least the same amount of information \cite{Reinagel1999}.
Bursts also have a greater ability to elicit responses on postsynaptic neurons \cite{Swadlow2001,CSICSVARI1998179}.%

Networks with the HB model have already had their synchronization characteristics studied in detail for different coupling strengths or temperatures in a small-world and scale-free topologies \cite{budzinski_detection,thiago_2014,budzinski2019temperature,batista_2013,boaretto2019protocol,xu2018synchronization,boaretto2019suppression,boaretto2018anomalous}. A transition from desynchronization, for very low coupling, to chaotic burst synchronization, for sufficiently strong coupling, was generally observed. The specific behavior between these two cases, however, was seen to differ strongly for different temperatures \cite{budzinski2019temperature,boaretto2018anomalous}.

For some values of temperature, in the coupling region before the transition to chaotic burst synchronization, the network was observed to be always desynchronized \cite{budzinski2019temperature,boaretto2018anomalous}. 
Increasing the temperature, this always desynchronized region changes to one with a local maximum in the synchronization quantifier, meaning that with the increase in coupling the network changes from desynchronized to burst synchronized and then to desynchronized again before the chaotic burst synchronization \cite{budzinski_detection,boaretto2019protocol}.
Further increase of the temperature makes the network start to synchronize even for very low coupling strengths. For some values of coupling, the synchronization is so strong that even the spikes within bursts are synchronized \cite{budzinski2019temperature}.
That is, three temperature intervals were observed for which networks have distinct synchronization behavior before an eventual transition to chaotic burst synchronization. 

Studying the uncoupled neurons, it was observed \cite{budzinski2019temperature} that there were also three different behaviors: for lower temperatures, neurons have various Inter-Burst Interval ($IBI$) values, indicative of chaotic behavior. With an increase of temperature, neurons then have two $IBI$ values. A further increase leads to only one $IBI$ value. 
The three synchronization behaviors for weakly coupled networks are related to the three $IBI$ behaviors for uncoupled neurons: for weak coupling, networks with temperature values corresponding to chaotic behavior in uncoupled neurons did not synchronize. In this same regime, for temperatures in which uncoupled neurons had two $IBI$s the coupled network had the local maximum in synchronization, as described previously. Still for the same coupling strengths, for temperatures in which uncoupled neurons had one $IBI$ the network synchronized strongly.

Therefore, it was concluded that, for these coupling strengths, the synchronization behavior of the coupled networks were linked to the local dynamics of the uncoupled neurons \cite{budzinski2019temperature}.

In the strong coupling regime, where networks achieved chaotic phase synchronization in a very similar way for all temperatures investigated, it was concluded that the forcing between neurons is strong enough to synchronize the network regardless of what the uncoupled dynamics is.
This also leads to the remark that the individual dynamics only plays an important role for weak coupling.
Applying a pulsed current in the network, the burst synchronization for weak coupling in the second temperature interval was successfully suppressed, while still maintaing the asymptotic synchronization \cite{boaretto2019protocol}.

In this work, we extend some of these results and show that the variability, measured by the standard deviation of the Inter-Burst Intervals ($IBI$) divided by their mean value, correlates with the degree of synchronization, measured by the Kuramoto Order Parameter \cite{kuramoto2012chemical}, for all the temperature values corresponding to the different behaviors mentioned previously.
We also show that there is a phenomenon we call promiscuity, which is the tendency of the relative phases between different neurons to change in time. The phases are calculated from the start of each neuron's firing (in the HB case, their bursts) and the relative phase is just the difference between the phases of different neurons. In this case, if promiscuity is zero, neurons are phase-locked and the higher the promiscuity the less neurons stay in phase with each other.

Promiscuity is shown to be different for network states with similar degrees of synchronization, but different coupling strengths - therefore allowing one to distinguish between them. 
We measure promiscuity in two different ways. 
The first is by counting how many neurons leave and enter the network cluster - defined here as all neurons within one standard deviation of the mean value of each bursting event.  

The second way calculates how much, on average, the burst start times of different neurons drift in relation to each other. The higher this value, the higher the probability that the difference between neurons' bursts times changes over time. This average drift method has the advantage of being parameter-free, corroborates the results obtained by the clustering analysis and can be easily applied to any time series of inter-burst or inter-spike intervals. 

The results obtained were for both small-world and random networks and are very similar.

We finally argue that variability generates promiscuity, due to a simple statistical process, described in the conclusions.

The paper is organized as follows: in section \ref{sec:methods} we introduce the methods we use - including the neuronal model and the synchronization quantifiers -, we also define the two types of variability, the clustering algorithm and the average drift measure. In section \ref{sec:results} we show the results and analysis, presenting, finally, the conclusions in section \ref{sec:conclusions}.

\section{Methods}
\label{sec:methods}

\subsection{Neuronal Model}
\label{sec:HB}
To simulate the neurons, we used an adapted Hodgkin-Huxley model, proposed by Braun \textit{et al} \cite{braun1998computer}. It consists of six differential equations and introduces the influence of temperature. The membrane potencial for the ith neuron at time $t$, $V_i(t)$, is given by the master equation
\begin{equation}
    C_\mathrm{M} \frac{dV_i(t)}{dt} = -J_{i, \mathrm{Na}} - J_{i, \mathrm{K}} - J_{i, \mathrm{sd}} - J_{i, \mathrm{sr}} - J_{i, \mathrm{L}} - J_{i, \mathrm{coup}},
\end{equation}
in which $C_\mathrm{M}$ is the specific membrane capacitance; $J_{i,\mathrm{Na}}$, $J_{i,\mathrm{K}}$, and $J_{i,\mathrm{L}}$ are the Sodium, Potassium, and Leakage current densities, respectively; $J_{i,\mathrm{sd}}$, $J_{i,\mathrm{sr}}$  are current densities related to the subthreshold depolarization (sd) and repolarization (sr); $J_{i, \mathrm{coup}}$ is the current density due to the interneuronal coupling.

The specific current densities $J_\alpha$, for $\alpha =  \{\mathrm{Na, K, sd, sr, L} \}$, are described by the Nernst potential $E_\alpha$ of the respective ion or leak channels:
\begin{align}
J_{i,\mathrm{Na}} &= \rho \bar{g}_\mathrm{Na}  a_{i,\mathrm{Na}} (V_i - E_\mathrm{Na}), \\
J_{i,\mathrm{K}}  &= \rho  \bar{g}_\mathrm{K}     a_{i,\mathrm{K}} (V_i - E_\mathrm{K}), \\
J_{i,\mathrm{sd}} &= \rho \bar{g}_\mathrm{sd}  a_{i,\mathrm{sd}} (V_i - E_\mathrm{sd}), \\
J_{i,\mathrm{sa}} &= \rho \bar{g}_\mathrm{sa}  a_{i,\mathrm{sa}} (V_i - E_\mathrm{sa}), \\
J_{i,\mathrm{L}}  &=  \bar{g}_\mathrm{L}         (V_i - E_\mathrm{L}).
\end{align}
The $\overline{g}_\alpha$ are maximum specific conductances, $\rho$ is the first scale factor, which serves to introduce the temperature dependence, and is given by
\begin{equation}
\label{eq:rho}
\rho = \rho_0^{(T-T_0)/\overline{T}_0}.
\end{equation}

The $a_\alpha$ are variables responsible for the channel activations, whose temporal evolutions are
\begin{align}
\frac{da_{i,\mathrm{Na}}}{dt} &= \frac{\phi}{\tau_\mathrm{Na}} (a_{i, \mathrm{Na}, \infty} - a_{i,\mathrm{Na}}), \\
\frac{da_{i,\mathrm{K}}}{dt}  &= \frac{\phi}{\tau_\mathrm{K}} (a_{i,\mathrm{K}, \infty} - a_{i,\mathrm{K}}), \\
\frac{da_{i,\mathrm{sd}}}{dt} &= \frac{\phi}{\tau_\mathrm{sd}} (a_{i,\mathrm{sd}, \infty} - a_{i,\mathrm{sd}} ), \\
\frac{da_{i,\mathrm{sa}}}{dt} &= \frac{\phi}{\tau_\mathrm{sa}} (-\eta J_{i,\mathrm{sd}} - \gamma a_{i,\mathrm{sa}})    
\end{align}
where the parameters $\eta$ and $\gamma$ describe the increase and decrease, respectively, of the concentration of $Ca^{2+}$ ions. The second scale factor $\phi$ is
\begin{equation}
\label{eq:phi}
\phi = \phi_0^{(T-T_0)/\overline{T}_0}.
\end{equation}
The $a_{i, \alpha, \infty}$ ($\alpha = \{ \mathrm{Na, K, sd}\}$) are activation functions which depend on the membrane potential through the equations
\begin{align}
a_{\mathrm{Na}, \infty} &= \frac{1}{1 + \exp[-s_\mathrm{Na}(V_i - V_{0\mathrm{Na}})]}, \\
a_{\mathrm{K}, \infty} &= \frac{1}{1 + \exp[-s_\mathrm{K} (V_i - V_\mathrm{0K})]}, \\
a_{\mathrm{sd}, \infty} &= \frac{1}{1 + \exp[-s_\mathrm{sd}(V_i - V_{0\mathrm{sd}})]}.
\end{align}

All previously unmentioned terms are constants, given in table \ref{tab:hb_csts}.
Finally, the coupling current $J_{i,\mathrm{coup}}$ is
\begin{equation}
  J_{i, \mathrm{coup}} = g_c\frac{\varepsilon}{\nu}(V_i - E_\mathrm{syn}) \sum\limits_{j \in \Gamma_i} r_j(t),
\end{equation}
  where $\varepsilon$ is the coupling parameter, which controls the strength of the coupling; $\nu$ is the normalization factor, defined as the maximum degree of connectivity of the network; $g_c$ is a unitary parameter with conductance units; $E_\mathrm{syn}$ is the synaptic reversal potential, whose value is taken to ensure that the synapse is always excitatory; $\Gamma _i$ is the set of neurons connected to the $i$-th neuron; $r_j$ is the coupling variable and has a temporal evolution given by \cite{destexhe1994efficient}
\begin{equation}
\frac{dr_j}{dt} = \left(\frac{1}{\tau_r} - \frac{1}{\tau_d}\right) \frac{1 - r_j}{1 + \exp[-s_0(V_j - V_0)]} - \frac{r_j}{\tau_d}.
\end{equation}
We note that the coupling current can be written as $J_{\mathrm{coup}} = \bar{g} P (V - E_\mathrm{syn})$ \cite{Dayan2005}, so that we can identify $\bar{g} = g_c\varepsilon/\nu$ as the maximum conductivity of the postsynaptic membrane and the summation as the fraction of bound postsynaptic receptors $P$. Thus, $r_j$ can be seen as the fraction of bound receptors of the $i$-th neuron that can be activated by the $j$-th neuron.

The integration is done using the CVODE solver \cite{hindmarsh2005sundials}, with a 12th order Adams-Moulton predictor-corrector method, absolute and relative tolerances of $10^{-6}$, and maximum time step $h = \SI{0.01}{\milli\second}$. The initial conditions are given randomly in the interval $[-65, 0]$ for $V$ and $[0,1]$ for the other variables. Simulations are run for $\SI{25}{\second}$, $\SI{15} {\second}$ of which are considered to be transient dynamics and disregarded.

The parameter values were obtained from \cite{braun1998computer}, with two alterations done for convenience, which do not alter the dynamical behavior: (i) temperatures were rescaled, by changing $T_0$, so that the relevant behavior lies in a range plausible for mammals (around $\SI{37}{\celsius}$); (ii) characteristic times were rescaled so that the frequency of bursts was around $\SI{10}{\hertz}$. Since the system is invariant under time translations, this change only affects the values of the burst times: except for these values, the dynamic behavior is exactly the same.

\begin{table*}
\caption{\label{table1}Parameter values of the constants for the Huber-Braun neuron model \cite{braun1998computer}. }
\resizebox{\linewidth}{!}{
\begin{tabular}{l l l l l}
  \hline\hline
  Membrane capacitance & \multicolumn{3}{c} {$C_{\mathrm{m}} = \SI{1.0}{\micro\farad/\centi\meter^{2}}$} \\ \hline
  Maximum conductances ($\SI{}{\milli\siemens / \centi\meter^{2}}$)\hspace{0.5cm} & $\overline{g}_{\mathrm{Na}}=1.5$  \hspace{1.5cm} & $\overline{g}_{\mathrm{K}}=2.0$  \hspace{1.5cm}  & $\overline{g}_{\mathrm{sd}}=0.25$ \hspace{1.5cm} & $\overline{g}_{\mathrm{sa}}=0.4$  \\
                       & $\overline{g}_{\mathrm{l}}=0.1$  &  $g_c \equiv 1.0$& &  \\ \hline
  Characteristic times ($\SI{}{\milli\second}$): & $\tau_{\mathrm{Na}}=0.05$ & $\tau_{\mathrm{K}}=2.0$ & $\tau_{\mathrm{sd}}=10$ & $\tau_{\mathrm{sa}}=20$  \\
  & $\tau_{\mathrm{r}} = 0.5$ & $\tau_{\mathrm{d}} = 8.0$\\ \hline
  Reversal potentials ($\SI{}{\milli\volt}$): & $E_{\mathrm{Na}}=50$   & $E_{\mathrm{K}}=-90$  & $E_{\mathrm{sd}}=50$  & $E_{\mathrm{sa}}=-90$  \\
                       & $E_{\mathrm{l}}=-60$  & $V_{0\mathrm{Na}}=-25$   & $V_{0\mathrm{K}}=-25$   & $V_{0\mathrm{sd}}=-40$ \\
                       & $E_{\mathrm{syn}} = 20 $    \\
  \hline
  Other parameters:  & $\rho_{0}=1.3$  & $\phi_{0}=3.0$  \\  
                       & $T_{0}=\SI{50}{\celsius}$ & $\overline{T}_{0}=\SI{10}{\celsius}$& $s_{\mathrm{Na}}=\SI{0.25}{\milli\volt^{-1}}$ & $\eta =\SI{0.012}{\centi\meter\squared/\micro\ampere}$   \\
                       & $s_{\mathrm{sd}}=\SI{0.09}{\milli\volt^{-1}}$ &   $\gamma=0.17$   & $s_{\mathrm{K}}=\SI{0.25}{\milli\volt^{-1}}$ & $s_0 = \SI{1.0}{\milli\volt^{-1}}$ \\
  \hline
 Network parameters: &$N = 1024$ & $\mathscr{N} = 7328$  & $\nu = 15$ \\
  \hline\hline
  \label{tab:hb_csts}
\end{tabular}%
}
\end{table*}

\begin{figure}[htb] 
\includegraphics[width=\columnwidth]{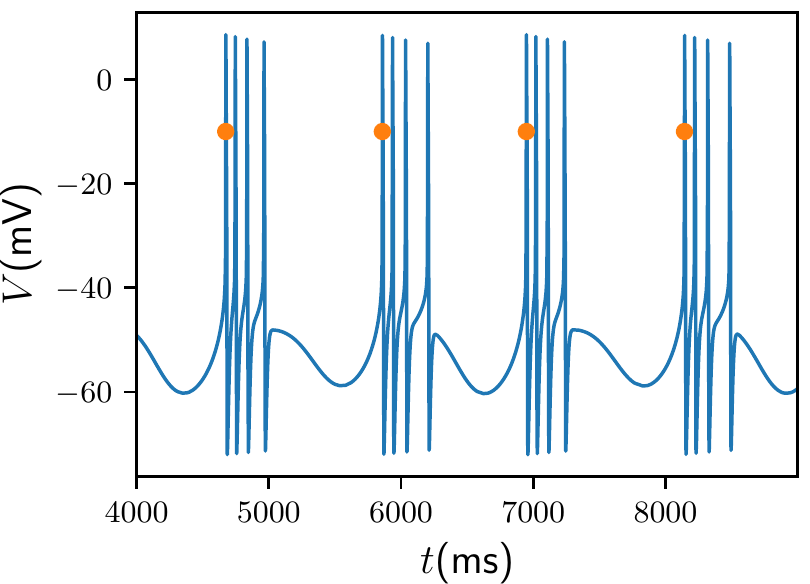}
\caption{Membrane potential $V$ of an uncoupled neuron, represented as the blue line, for $T = \SI{38}{\celsius}$ and the burst start times, represented as the orange circles.}
\label{fig:hb-V}
\end{figure}

In \figref{fig:hb-V} we show a typical membrane potential (blue line) for an uncoupled neuron and the corresponding burst start times (orange circles). 
 
 \subsection{Topologies}
We used the small-world topology proposed by Newman and Watts \cite{newman_1999}, in which a regular network of size $N$ and $k = 4$ neighbors is generated and then a number $\mathscr{N} - 2kN$ of connections is added, thus totalling $\mathscr{N}$ connections. This number of added connections is chosen in such a way that the average path length of the network is low (comparable to a random topology) and the clustering coefficient is high (much higher than random), so that this network is locally clusterized, but still neurons are not so far away (in a number of connections sense) from each other \cite{watts_1998}. 

The network generated with this algorithm is directed, meaning that if neuron $i$ is connected $j$, then $j$ is not necessarily connected to $i$. This is chosen to be so because the synapses are chemical and therefore not symmetrical. The network also has no self-loops, since we consider that neurons don't connect back to themselves. 

Another algorithm, proposed by Watts and Strogatz \cite{watts_1998} also has been used to generate the small-world networks, with similar results being obtained.

\subsection{Synchronization Quantifiers}
\label{sec:quantifiers}
Phase synchronization is a state in which the oscillators of the system have similar phases, but may have different amplitudes \cite{synchronization_universal_book}. To quantify the degree of synchronization of a neural network, we first introduce the phase $\theta_i(t)$ for each neuron. We define it as starting in $0$, increasing by $2\pi$ every time a burst starts and being a linear interpolation in between \cite{ivanchenko_2004}
\begin{equation}
\theta_{i}(t) = 2\pi k_i + 2\pi \frac{t - t_{k,i}}{t_{k+1, i} - t_{k,i}}, \; (t_k < t < t_{k+1})
\end{equation}
where $t_{k,i}$ is the time at which the $k$-th burst of the $i$-th neuron ocurred, called the burst start time.

We remark that this definition works for all parameter values studied in this work. For coupling strengths higher than the ones used, single spikes start appearing a significant distance between adjacent bursts and so a problem arises when defining a phase. In the cases we studied, however, this is not significant because these events are rare and the distances are small, so the isolated spike are considered to belong to the previous burst.

The measurement of the phase synchronization is done via the Kuramoto order parameter \cite{kuramoto2012chemical}
\begin{equation}
\label{eq:kuramoto}
R(t) = \frac{1}{N}\left|\sum\limits_{j=1}^N e^{i \theta_j(t)}\right|,
\end{equation}
where $i$ is the imaginary unit here. The quantifier $R$ is $0$ if the network is completely desynchronized (each neuron has another neuron with opposite phase) and is $1$ if the network is completely synchronized in phase (every neuron has the same phase).
We may take the time average,
\begin{equation}
\label{eq:kuramoto_mean}
  \langle R \rangle = \frac{1}{n} \sum_{t=t_0}^{t_f} R(t),
\end{equation}
    where $t_0$ is the transient time and $t_f$ is the total simulation time and $n = (t_f - t_0)/h$ is the number of steps, with $h$ being the time step. 

\subsection{Variability}
\label{ssec:variability}
We define the $k$-th Inter-Burst Interval (IBI) of the $i$-th neuron as the difference between its $k$th and $(k+1)$th burst start times 
\begin{equation}
  IBI_{k,i} = t_{k+1,i} - t_{k,i}.
\end{equation}

The variability is measured by the coeficient of variability $CV$, defined as \cite{Softky334, stevens1998}
\begin{equation}
\label{eq:cv}
  CV = \frac{\sigma (IBI) }{ \langle IBI \rangle },
\end{equation}
where $\langle IBI \rangle$ is the time average taken over the IBIs of all neurons and $\sigma(IBI)$ is the standard deviation of the IBIs, which may be calculated in 2 different ways: (1) by taking the spatial average of the deviation of a neurons' IBI over time, denoted $CV_t$; (2) by taking the time average of the deviation over space, denoted $CV_s$: 
\begin{align}
    \sigma_t &= \frac{1}{N}       \sum_{i=1}^{N} \sigma \bigg(\sum_{k=1}^{k_{\mathrm{max}}} IBI_{k,i}\bigg) \\
    \sigma_s &= \frac{1}{k_{\mathrm{max}}} \sum_{k=1}^{k_{\mathrm{max}}}\sigma \bigg(\sum_{i=1}^{N} IBI_{k,i}\bigg), 
\end{align}
in which $k_{\mathrm{max}}$ denotes the total number of $IBI$s analyzed.

\subsection{Clustering Analysis}
\label{sec:cluster_analysis}

\begin{figure}[hbt] 
\includegraphics[width=\columnwidth]{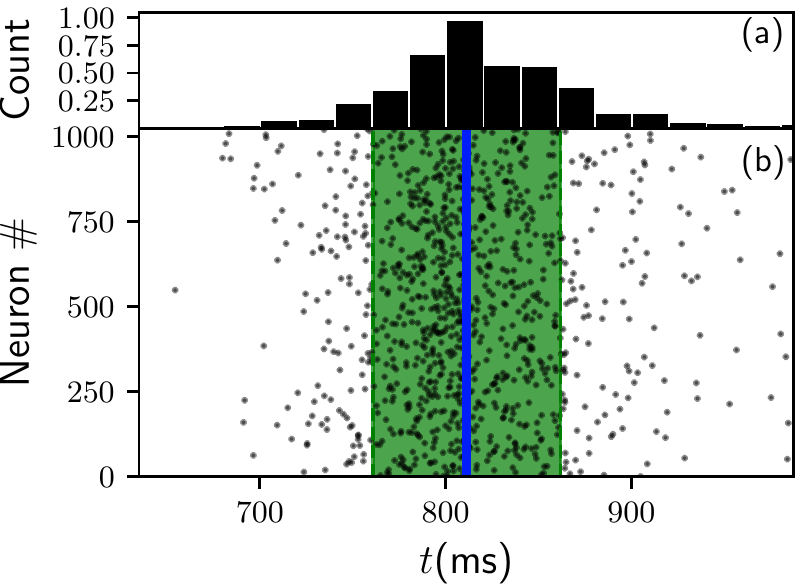}
\caption{Panel (b) depicts the network raster plot for $T = \SI{38}{\celsius}$ and $\varepsilon = 0.00879$. Here, each black circle represents the time where a burst starts. The blue line represents the mean value of the bursting start times and the green area the standard deviation, which defines the cluster. Panel (a) shows the normalized histogram from the bursting start times depicted in panel (b).}
\label{fig:RP_def}
\end{figure}

In a phase synchronized network, neurons' burst times tend to be distributed closely around a mean value for every bursting event. In panel (b) of \figref{fig:RP_def} are depicted the burst start times of all neurons in a network for one single event for $T = \SI{38}{\celsius}$ and $\varepsilon = 0.00879$. From the corresponding histogram (panel (a)), one can see that neurons tend to group themselves around a mean value.

To quantify this observation, we first define a cluster to be the set of neurons that are within one standard deviation from the mean value of the bursting event. In \figref{fig:RP_def}, the mean value corresponds to the blue line, and the cluster to the green area.

We are interested in studying how the relative bursting times of neurons change with time. In this way, it is appropriate to study the changes in the cluster composition throughout time. 
With this in mind, we generalize the cluster definition to consider the behavior for multiple bursting events in the following way: identify the clusters $C_k$ for every bursting event $k$ for $\mathcal{T}$ events and consider only neurons who are in all clusters. This intersection $C_1 \cap C_2 \cap \dots \cap C_\mathcal{T}$ will be called a $\mathcal{T}$-cluster and the collection of $\mathcal{T}$ bursting events is called a cluster event.

This can be summarized in the following algorithm
\begin{enumerate}
    \item Identify the approximate time of the bursting event through a local maximum on the histogram of the burst times. 
    \item Choose the burst times that are sufficiently close (within one half of a bursting period) for each neuron.
    \item Take the mean value of the burst times obtained above, resulting in the mean burst time.
    \item Collect all neurons whose burst time is within one standard deviation from the mean burst time in a cluster. 
    \item Repeat the above for $\mathcal{T}$ events.
    \item Take the intersection of all the clusters obtained above to get the $\mathcal{T}$-cluster.
\end{enumerate}

The number of neurons inside each cluster $C_k$ is naturally defined as the cluster size $CS_k$.
Once all clusters are identified, we select two adjacent clusters, described by the sets $C_k$ and $C_{k+1}$, respectively, and count the number $\mathcal{L}_k$ of neurons that leave the cluster (i.e. the size of the set $C_k \setminus C_{k+1}$).
Finally, taking the average over various bursting events, we obtain the mean cluster size $CS$ and the mean number of neurons that leave the clusters $\mathcal{L}$.
A downside of this clustering analysis is its necessity to identify what we call a bursting event and identifying which of a neuron's burst start times correspond to this event. This relies on a certain minimum degree of synchronization of the network: if the network is unsufficiently synchronized, the analysis cannot be applied. In that case, another method needs to be considered. We introduce such a method in the next subsection (\ref{ssec:drift}).

\subsection{Average temporal drift}
\label{ssec:drift}
In this section, we define a quantity to measure if the temporal distances between neurons' bursts remain locked or change in time. To do this, we start with the time $t_{k,i}$ of the start of $k$-th burst of the $i$-th neuron, and then calculate the temporal distance of each pair of neurons $(i,j)$ for each event $k$
\begin{equation}
    \delta_{ij}^k = \lvert t_{k,i} - t_{k,j} \rvert.
\end{equation}

Then we calculate the absolute difference of this distance between sucessive events for each pair $(i,j)$,
\begin{equation}
    \Delta_{ij}^l = \lvert \delta_{ij}^k - \delta_{ij}^{k-1} \rvert.
\end{equation}

The temporal average of $\Delta_{ij}$, $\langle \Delta_{ij} \rangle$, measures the tendency of the pair $(i,j)$ to drift away from each other across time. 
We average the result over all pairs of neurons, resulting in $\Delta \equiv \overline{\langle \Delta_{ij} \rangle}$, which is termed the average drift of the network.
This average drift tendency $\Delta$ measures how much, on average, the temporal distances between neurons' firings change. 
If it is low, neurons are locked together, the difference between their burst start times remaining fixed. If it is high, neurons are not phase-locked.

\section{Results and discussions}
\label{sec:results}

\subsection{Variability and degree of synchronization}

\begin{figure*}[hbt] 
    \centering
    \includegraphics[clip,width=0.98\textwidth]{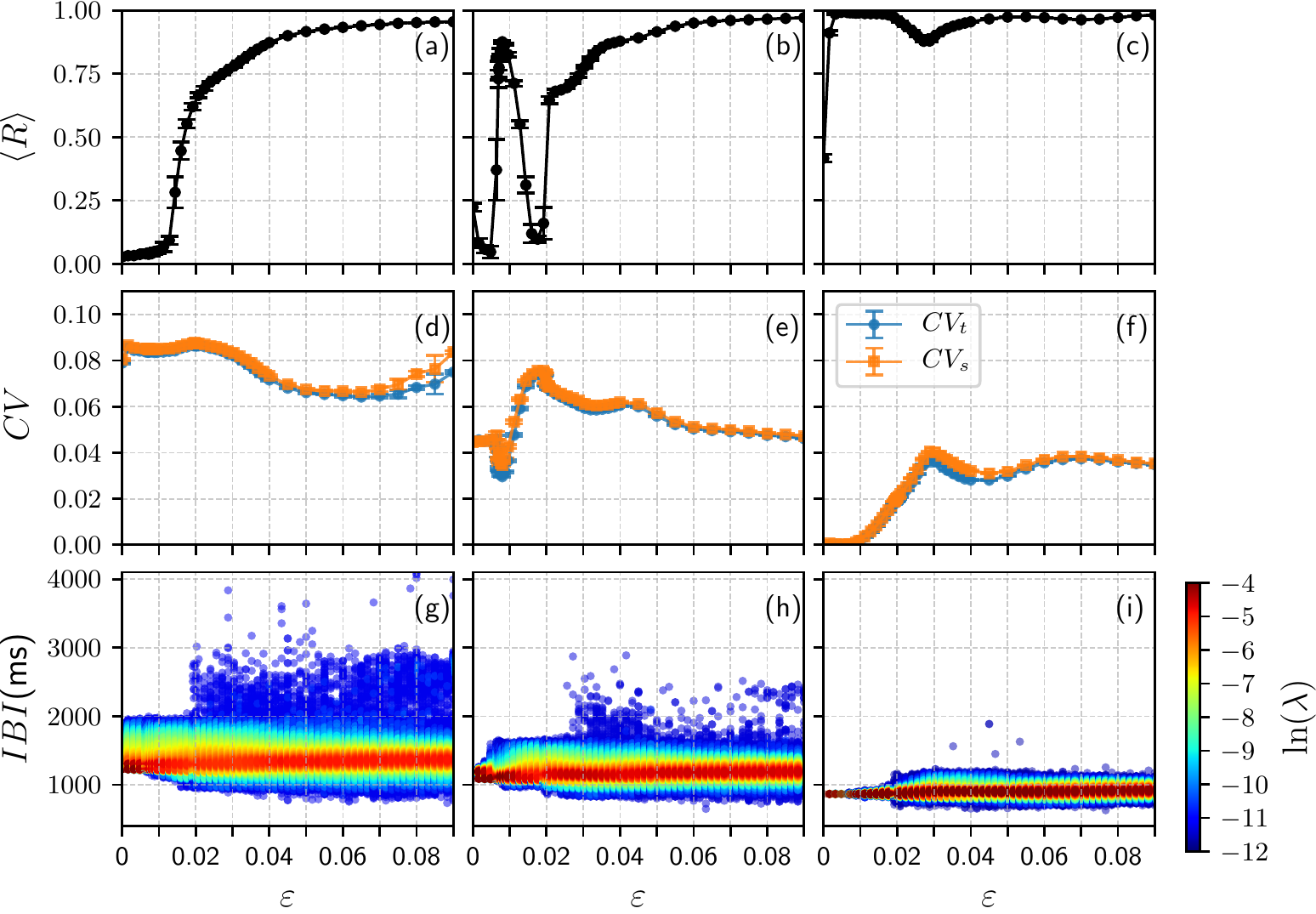}
    \caption{The scenario of synchronization and its relation to the neurons' variability as a function of coupling strength for different values of temperature. 
    The first row corresponds to the mean Kuramoto order parameter ($\langle R \rangle$); the second to the coefficients of variability ($CV_t$ and $CV_s$); the third to the $IBI$ bifurcation diagram. The columns correspond, from left to right, to temperatures $T = \SI{37}{\celsius}$,  $T = \SI{38}{\celsius}$, and $\SI{40}{\celsius}$.
    Different behaviors are observed: in the first column ($T = \SI{37}{\celsius}$), a traditional transition from non-synchronized to a phase-synchronized state is observed, with variabilities being high. Increasing the temperature to $T = \SI{38}{\celsius}$ and $T = \SI{40}{\celsius}$, in the second and third columns, leads the system to a different scenario, with a non-monotonic synchronization level phenomena. In these cases, the local maxima of $\langle R \rangle$ happens just where variabilities ($CV_{t}$ and $CV_{s}$) are low. Results are given by an average over $10$ initial conditions.}   
    \label{fig:R-var}
\end{figure*}

The main scenario of synchronization and the neurons' variability as a function of the coupling strength $\varepsilon$ is depicted in \figref{fig:R-var}. The degree of synchronization, measured by the time-averaged Kuramoto order parameter (Eq. \ref{eq:kuramoto_mean}), is shown in the first row of \figref{fig:R-var} for different values of temperature $T$. 

The coupling strength values are in the interval $[0, 0.09]$, with the uncoupled case being the first value used.

Panel (a) depicts the behavior for $T = \SI{37}{\celsius}$, in which a monotonic transition to phase synchronization is observed, similar to that noticed in Kuramoto Oscillators \cite{kuramoto2012chemical}. In panel (b), for $T = \SI{38}{\celsius}$, there is a local maximum of the synchronization quantifier in the weak coupling regime, characterizing a non-monotonic transition \cite{budzinski_detection}. Panel (c) corresponds to $T = \SI{40}{\celsius}$. In this case, the local maximum of $\langle R \rangle$ is replaced by a global maximum, in which spike synchronization can be observed for weak coupling, as reported in \cite{budzinski2019temperature}.

We notice that the network has very different behaviors for weak coupling, ranging from desynchronization to burst synchronization to even complete synchronization by a change of temperature $T$. For high coupling strengths, roughly in $[0.06,0.09]$, the behavior is almost independent of $T$, due to the network having chaotic phase synchronization \cite{budzinski2019temperature}.

The second row of \figref{fig:R-var} shows the variabilities, as defined in section \ref{ssec:variability}: temporal variability ($CV_{t}$) is in blue and spatial variability ($CV_{s}$) is in orange. Panel (d) ($T = \SI{37}{\celsius}$) depicts high values of $CV$ for small values of $\varepsilon$, which decreases in the high coupling regime. Panel (e) ($T = \SI{38}{\celsius}$) depicts a $CV$ minimum for the weak coupling regime, which is followed by a maximum. Increasing the coupling further, $CV$ decreases, but not below the previous minimum. 
In panel (f) ($T = \SI{40}{\celsius}$), we see a significant decrease in the two variabilities. Following the periodic behavior in the uncoupled case, the variabilties are very close to zero for very weak coupling ($\varepsilon \lesssim 0.01$). After that, the $CV$s start to increase and a corresponding decrease in the degree of phase synchronization $\langle R \rangle$ is observed in panel (c). This culminates in a maximum of the $CV$s, which happens together with a minimum in $\langle R \rangle$.

These results show a clear correlation for the weak coupling regime ($\varepsilon \lesssim 0.02$): a high variability ($CV_{t}$ or $CV_{s}$) is associated with low degree of synchronization ($\langle R \rangle$), as in $T = \SI{37}{\celsius}$ and low variability is associated with high synchronization (for both $T = \SI{38}{\celsius}$ and $T = \SI{40}{\celsius}$).
This relation is especially clear for the last two temperatures, in which regions with almost zero variability are the ones with almost complete synchronization.

The third row of \figref{fig:R-var} shows the bifurcation diagrams for the Inter-Burst Intervals of neurons in the network. 
The color of each point is determined by the logarithmic density $\ln(\lambda)$ of the corresponding $IBI$. The density $\lambda$ is equal to the number of times the $IBI$ was observed divided by the total number of $IBI$s observed. The colorbar, shown on the right, has values in the interval $[-12, -4]$.
These bifurcation diagrams explain most of the transitions in the dynamics of the network when the coupling strength is varied. Those transitions are reflected by characteristic changes in the probability distribution of $IBI$s, particularly in the width of the probability density function, leading to larger standard deviations or in the location of its maximum value leading to a shift in the mean value. The changes in variability are related to changes in the width and each of the transitions described above in the slope of variability can be associated to such changes in the width of the $IBI$ PDF.

An important detail consists in the behavior for temperatures $T = \SI{38}{\celsius},\;\SI{40}{\celsius}$, in which the network desynchronizes as coupling strength is increased. 
The analysis suggests that, in these cases, the coupling strength begins to be strong enough to induce a more chaotic behavior (reflected in the increase of the variability), which desynchronizes the network. 

A further increase in the coupling strength makes the network assume a chaotic burst synchronization.
This result is in line with previous observations \cite{ budzinski2019temperature,boaretto2018anomalous} that the local, individual behavior of the neurons is very relevant to the global, coupled behavior for weak coupling. For higher coupling strengths that is no longer the case, since forcing is strong enough to be dominant. Here, we arrive at this conclusion from the variability statistical analysis, a different approach than the one used previously: instead of using the individual membrane potentials or the mean field, we analyse just the burst times.

\subsection{Promiscuity}
\label{ssec:promiscuity}
    \subsubsection{Raster Plots}

\begin{figure}[htb] 
\centering
\includegraphics[width=0.98\columnwidth]{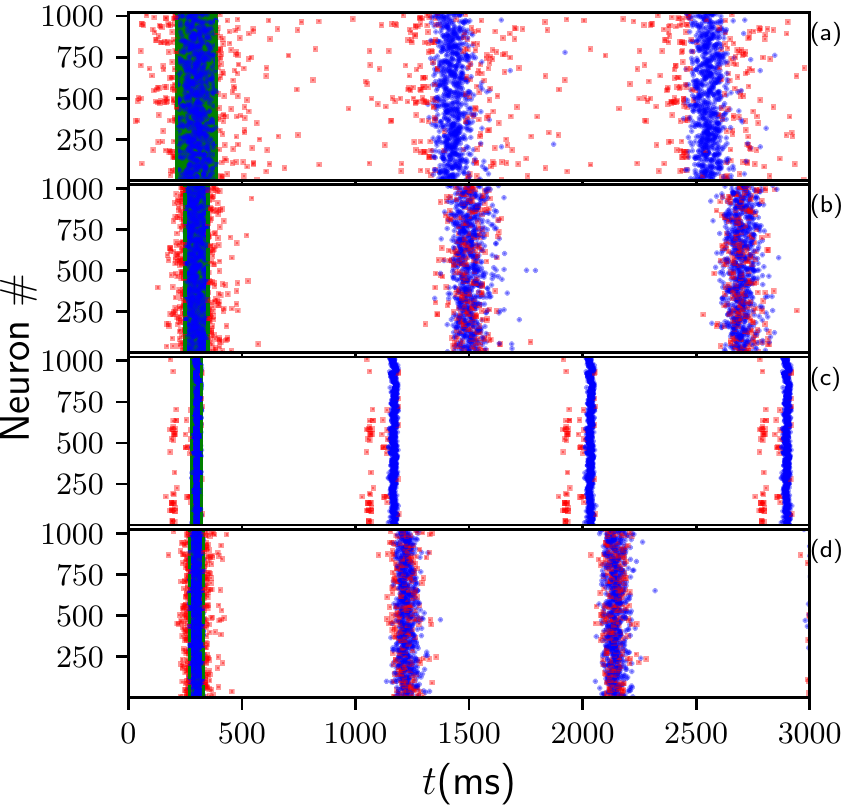}
\caption{Raster plots of the burst start times for different values of temperature $T$ and coupling strength $\varepsilon$. Blue points correspond to neurons that were in the first cluster (defined with $\mathcal{T} = 1$) of the graph and red squares to neurons that were not. The region painted green displays the interval within which neurons are considered to be in the cluster, which is defined by the standard deviation of the bursting start times. Here, panels (a) and (b) correspond to the case of $T = \SI{38}{\celsius}$ where $\varepsilon = 0.00879 $ and $\varepsilon = 0.09 $, respectively. Panels (c) and (d) represent the case of $T = \SI{40}{\celsius}$ for $\varepsilon = 0.00879 $ and $\varepsilon = 0.09 $, respectively. For both cases, in the region of smaller coupling strength neurons tend to stay in the cluster, while for the higher coupling neurons tend to mingle more. Time $t$ shown is the time after $15 s$ of transient dynamics.}
\label{fig:rp}
\end{figure}

To get more information about the synchronization characteristics of the network, \figref{fig:rp} depicts the raster plot of the burst starting times for the synchronization states of the network in the weak and strong coupling regimes.  In these cases, after the transient time ($t_{0}$), we identify the cluster, using the protocol defined in section \ref{sec:cluster_analysis}, for the first bursting event. All neurons inside the first cluster (green) are painted blue and all outside are painted red. For subsequent events, this coloring scheme is mantained, which means that a blue (red) dot in \figref{fig:rp} represents a neuron that was (not) inside the cluster in the first event (green area). In this context, panel (a) depicts the network for $T = \SI{38}{\celsius}$ and $\varepsilon = 0.00879$, the local maximum in \figref{fig:R-var}. We note that neurons inside the cluster tend to remain inside, while those outside tend to remain outside. 
The opposite is true in panel (b), where the network has the same temperature but the coupling strength is increased to $\varepsilon = 0.09$ (a chaotic phase-synchronized state). In this case, there is very strong mingling between neurons (blue and red dots are mixed).

A similar behavior is observed for the network with $T = \SI{40}{\celsius}$. Panel (c) of \figref{fig:rp} depicts the state of synchronization in the weak coupling regime ($\varepsilon = 0.00879$), in which a mixture of blue and red dots is not observed. This phenomenon is very accentuated in this case: the profile seems frozen in time (i.e. there is almost zero promiscuity). This reflects the almost zero degree of variability.
Finally, panel (d) depicts the raster plot for $\varepsilon = 0.1$ and $T = \SI{40}{\celsius}$ where there is a mixture between the blue and red dots in a very similar way to the behavior of a network with the same coupling and $T = \SI{38}{\celsius}$. The difference in temperature does not seem to affect the behavior for the high coupling regime.

Given the definition of promiscuity, introduced in section \ref{ssec:promiscuity}, the cases stated previously in \figref{fig:rp}, $\varepsilon = 0.00879$ (panels (a) and (c)) have low promiscuity (neurons are more phase-locked) and $\varepsilon = 0.1$ (panels (b) and (d)) have high promiscuity (neurons are less phase-locked).

It is also noteworthy that the promiscuity follows the same tendencies as the spatial variability, shown in \figref{fig:R-var}.

\subsubsection{Clustering Analysis}
\begin{figure}[htb] 
  \centering
  \includegraphics[width=0.98\columnwidth]{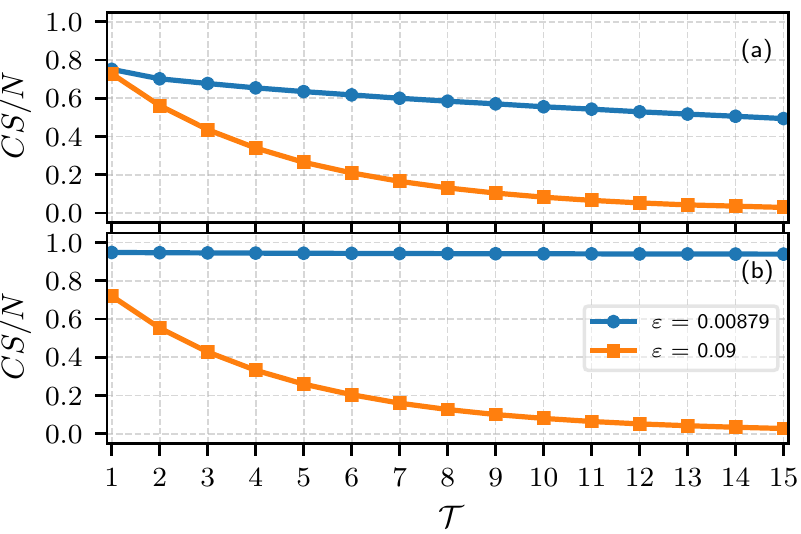}
  \caption{Cluster size $CS$ normalized by the network size $N$ as a function of the number of bursting events $\mathcal{T}$ for $T = \SI{38}{\celsius}$ (panel a) and $T = \SI{40}{\celsius}$ (panel b). The cluster is defined following section \ref{sec:cluster_analysis} and its size's dependence on $\mathcal{T}$ refletcs how many neurons remain together for $\mathcal{T}$ bursting events. Blues lines ($\varepsilon = 0.00879$) represent the synchronized states for the weak coupling regime, while oranges lines ($\varepsilon = 0.09000$) do so for the high coupling regime. The slope of the curve is proportional to the variability: orange lines, corresponding to high variability, have a steeper decline compared to the blue ones, with low variability. This is most visible in panel (b), in which the blue line (almost zero variability) has an almost  zero decline. Results are averaged over $10$ initial conditions, with vanishing errorbars not shown.
  }
  \label{fig:sizeClusterxt}
\end{figure}

The previous observations, depicted in \figref{fig:rp}, can be made more quantitative by identifying the $\mathcal{T}$-clusters (as defined in section \ref{sec:cluster_analysis}) for all clustering events and calculating $CS/N$ (the average cluster size $CS$ normalized by the network size $N$) as a function of the number $\mathcal{T}$ of bursting events. This is done in \figref{fig:sizeClusterxt}, in which panel (a) represents the behavior for $T = \SI{38}{\celsius}$ and panel (b) for $T = \SI{40}{\celsius}$. In both panels, blue lines correspond to $\varepsilon = 0.00879$, representative of the synchronized state for weak coupling, and orange lines correspond to $\varepsilon = 0.09$, representative of the strong coupling synchronization (see \figref{fig:R-var}).

The results of \figref{fig:sizeClusterxt} show a similar behavior of $CS/N$ for both temperatures considered: in the weak coupling regime (blue lines), the normalized size of the cluster tends to remain high even for high values of $\mathcal{T}$,  contrary to the strong coupling regime (orange lines), in which $CS/N$ tends to vanishing values as the number of bursting events is increased. These results indicate that neurons tend to stay in the cluster for longer times (i.e. promiscuity is lower) for the weak coupling cases. This phenomenon is especially clear for $T = \SI{40}{\celsius}$, in which the blue line has zero decline, indicating that all neurons in the $\mathcal{T}$-cluster stay together through at least $15$ subsequent events. 

It is worth remembering that networks in the strong coupling regime have the same or even higher values of the average Kuramoto order parameter $\langle R \rangle$, in comparison to ones with weak coupling. Even still, they have higher promiscuity, showing that higher degree of synchronization does not imply in longer duration of phase-locking (i.e. lower promiscuity).
The curves for strong coupling ($\varepsilon = 0.09$) are very similiar for the two temperatures, reflecting the fact that dynamics in this case is really due to the coupling, not to the local dynamics (which is affected by the temperatures) \cite{budzinski2019temperature}.

\begin{figure*}[htb] 
  \centering
  \includegraphics[width=\textwidth]{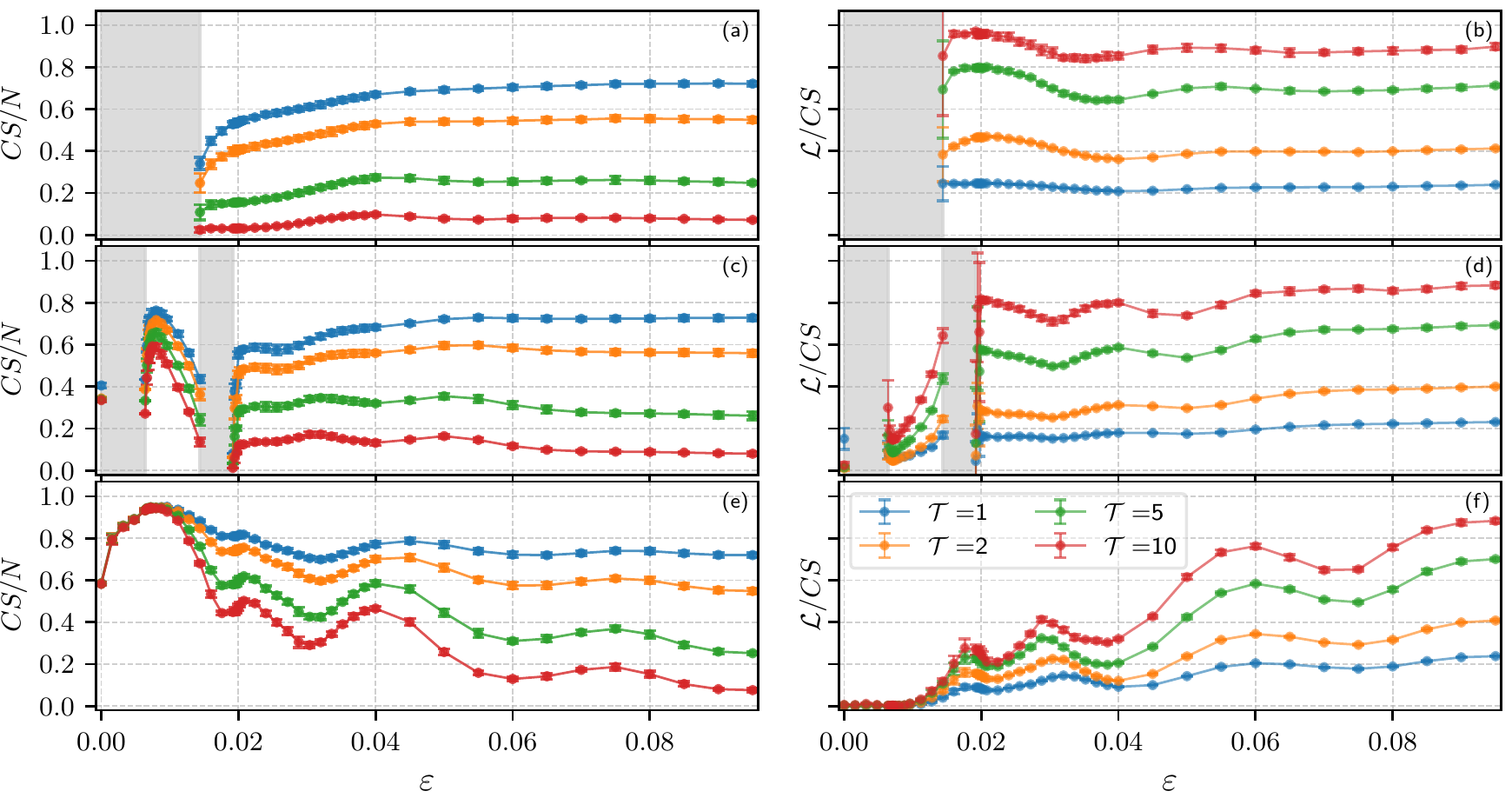}
   \caption{Normalized cluster size ($CS/N$) (first column) and normalized number of neurons that leave the clusters ($\mathcal{L}/CS$) (second column) as a function of coupling strength $\varepsilon$ for temperatures $T = \SI{37}{\celsius}$, $T = \SI{38}{\celsius}$, $T = \SI{40}{\celsius}$ (first, second and third rows, respectively). The line colors are varied according to the number $\mathcal{T}$ of events needed for neurons to be grouped in the cluster: blue lines are for $\mathcal{T} = 1$, orange for $\mathcal{T} = 2$, green for $\mathcal{T} = 5$, and red for $\mathcal{T} = 10$. Regions in which no cluster could be defined are depicted in light gray. $CS/N$ has a higher value, and smaller decrease as $\mathcal{T}$ increases, in the weak coupling regions, where spatial variability is smaller. The lower decrease in these regions indicates that neurons remain in the cluster for longer times. A similar information can be observed for $\mathcal{L}/CS$, in which the quantifier is lower for weak coupling, meaning that fewer neurons leave the cluster in this case. Also, the correlation between $\mathcal{L}/CS$ and the spatial variability is notable. All values are averages over $10$ different initial conditions, with the errorbars corresponding to the standard deviation over them.}
  \label{fig:numLeftxeps}
\end{figure*}

We extend the previous analysis for all coupling strengths in \figref{fig:numLeftxeps}, which depicts the normalized cluster size $CS/N$ (first column) and the average normalized number of neurons that leave subsequent clusters $\mathcal{L/CS}$ (second column) as a function of the coupling strength $\varepsilon$. Each row corresponds to different values of temperature: $T = \SI{37}{\celsius}, \SI{38}{\celsius}, \SI{40}{\celsius}$ from top to bottom, respectively. Furthermore, different values of the number of bursting events $\mathcal{T}$ are considered, represented in different line colors: blue for $\mathcal{T} = 1$, orange for $\mathcal{T} = 2$, green for $\mathcal{T} = 5$ and red for $\mathcal{T} = 10$.  The light gray areas are regions where it is not possible to define cluster by using the approach defined in the section \ref{sec:cluster_analysis}.

The difference in behavior is much more stark for weak coupling strengths ($\varepsilon < 0.04$). In this region, for $T = \SI{37}{\celsius}$ (panels (a) and (b)) there is no defined cluster since the network is not sufficiently synchronized. For $T = \SI{38}{\celsius}$ (panels (c)), there is a high number of neurons in the cluster, with a peak at the local maximum of the Kuramoto Order Parameter, shown in \figref{fig:R-var}. Increasing the number $\mathcal{T}$ of events, the normalized cluster size $CS/N$ decreases, but not by much, indicating that most neurons stay in the cluster for at least $\mathcal{T} = 10$ events, which is reflected in the low value of $\mathcal{L}/CS$ (panel (d)).
Still for weak coupling, for $T = \SI{40}{\celsius}$ (panel (e)) almost the whole network is clustered ($CS/N \approx 1$).
For these two temperatures, in the very weak coupling interval ($\varepsilon \lesssim 0.01$), there is almost no change in the value of $CS/N$ as $\mathcal{T}$ is increased, which is corroborated by the value of $\mathcal{L}/CS$, in panel (f) and is a quantitative representation of what was observed in the raster plot of \figref{fig:rp}. 
This region is precisely the one with almost zero spatial variability ($CV_s$) (cf. \figref{fig:R-var}).

For higher values of coupling, ($\varepsilon \gtrsim 0.04$), in $T = \SI{40}{\celsius}$, the decrease with $\mathcal{T}$ is much more significant, reflecting the increase of $CV_s$. 
In this regime the behavior is very similar for all temperatures: neurons group together around a cluster, but do not remain inside for more than, at most, a few bursting events: that is, the duration of phase locking is very short.

In short, comparing regions of weak coupling with ones of high coupling, we note that the degree of synchronization (measured by $\langle R \rangle$) may be similar, but the promiscuity is very different: networks with weaker coupling are less promiscuous in these cases.

Furthermore, we note that, once again, promiscuity and variability follow the same tendencies, as can be seen by comparing $CV_s$ in \figref{fig:R-var} with $\mathcal{L}/CS$ in \figref{fig:numLeftxeps}.

\subsubsection{Average temporal drift}

At last, a parameter-free approach is used to verify the results about the promiscuity of a neural network. As defined in the section \ref{ssec:drift} the average temporal drift ($\Delta$) measures the average changes in the distance between burst times.

\begin{figure}[htb] 
  \centering
  \includegraphics[width=\columnwidth]{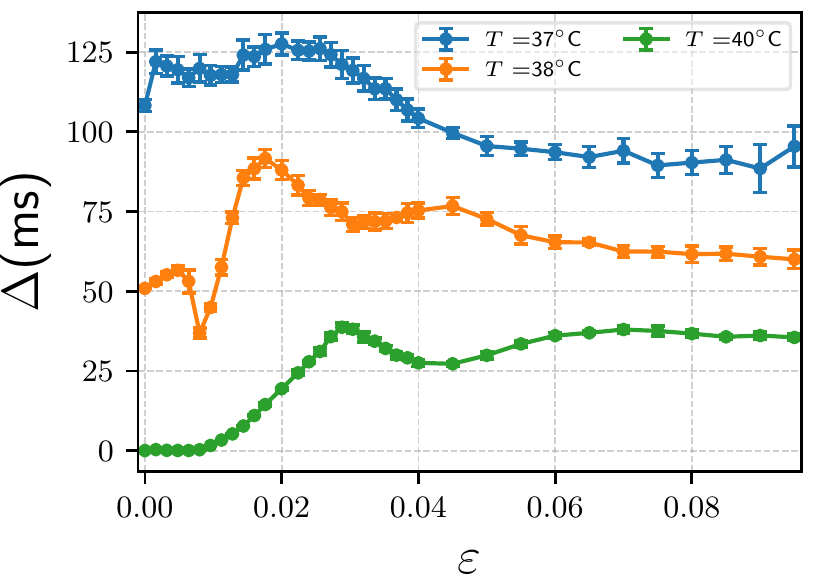}
  \caption{Average drift $\Delta$ as a function of the coupling strength $\varepsilon$. The blue, orange and green lines refer to $T = \SI{37}{\celsius}$, $T = \SI{38}{\celsius}$, and $T = \SI{40}{\celsius}$, respectively. 
    The average drift is a second measure of promiscuity and, as such, corroborates the results obtained via the clustering method in \figref{fig:numLeftxeps}. It has the advantages of being parameter-free and working for all degrees of synchronization in the network.   
Results are averaged over 10 initial conditions, with errorbars equal to the standard deviation over them.}
  \label{fig:drift}
\end{figure}

\figref{fig:drift} shows $\Delta$ as a function of the coupling strength $\varepsilon$ for different temperature, as considered in main scenario (\figref{fig:R-var}).

For $T = \SI{37}{\celsius}$ (blue line), the drift starts high and decreases with an increase of coupling, indicating higher promiscuity for weaker coupling strengths. This was not observed with the clustering method, as depicted in \figref{fig:numLeftxeps} since synchronization in this region is so weak that a cluster could not be defined.

For $T = \SI{38}{\celsius}$ (orange line), the average temporal drift $\Delta$ has a minimum in the region of weak coupling, coinciding with the local maximum of $\langle R \rangle$ (\figref{fig:R-var}). A coupling strength increase makes $\Delta$ go up until a maximum is reached, corresponding to the local minimum of $\langle R \rangle$.
Further increase of coupling decreases $\Delta$, but the final value is still higher than in the low coupling regime.

Similar phenomena are observed for $T = \SI{40}{\celsius}$ (green line): there is lower promiscuity (lower $\Delta$) for weak coupling ($\varepsilon \lesssim 0.04$), where spatial variability is also lower - and higher promiscuity (higher $\Delta$) for high coupling ($\varepsilon \gtrsim 0.04$), where spatial variability is higher.

Promiscuity (measured in this case by $\Delta$) has a non-monotonic dependence on the coupling strength ($\varepsilon$) and this dependence profile changes with the temperature $T$.
However, the relation between promiscuity $\Delta$ and the spatial variability $CV_s$, observed in \figref{fig:R-var} is maintained in all cases: the two quantities have the same tendencies. 
This coincidence of tendencies between the drift $\Delta$ and the spatial variability $CV_s$ calls into question if they are equivalent by construction. This is in fact not the case: it is possible for $\Delta$ to change, but for $CV_s$ to remain constant. 

Very similar results were also observed in random networks with the same number $\mathscr{N}$ of connections, created through the algorithm proposed in \cite{erdos1960evolution}.

\section{Conclusions}
\label{sec:conclusions}
The Inter-Burst Interval ($IBI$) variabilities were measured in a network of modified-Hodgkin-Huxley neurons and shown to affect the phase synchronization of the network. Firstly, for small coupling strengths, the degree of phase synchronization was observed to be inversely proportional to both temporal and spatial variabilities. This correlation is observed even for different behaviors: from no synchronization to almost complete synchronization.

Secondly, spatial variability was shown to be positively correlated with neuronal promiscuity, the tendency of neurons to change their relative phases in time.
That is, for low variability, the neuronal promiscuity is low: neurons tend to stay together throughout time; for high variability, promiscuity is high: neurons tend to mingle. 

It is striking that, in general, promiscuity is not directly related to the degree of synchronization (it was observed only for small coupling strengths). As such, there can be highly synchronized networks with low promiscuity, but also ones with high. In fact, this high promiscuity is present in all investigated temperatures for high coupling strengths, which is counterintuitive: in those cases, neurons that are strongly coupled are also less locked together.

Since, according to the communication-through-coherence hypothesis \cite{CNC_FRIES}, absence of phase coherence may prevent communication between neuronal groups, then we note that this relation between variability and phase-locking is important to keep in mind in the study of neural communication. 

We argue that the relation between spatial variability and promiscuity is actually causal, in that the former actually generates the latter. This occurs simply because the higher the variability, the higher the pool of available $IBI$s, so the higher the chance neurons will choose different $IBI$ and therefore the harder it is for neurons to stay locked together (ie the higher the promiscuity). This effect is statistical in nature, so it should be present to some extent in networks with spatial variability in the inter-firing (spikes or bursts) intervals. In fact, it was always observed for this neuronal model, with small-world and random topologies, as shown in the paper and in the appendix, respectively.

It is worth noting that some regions of different coupling strengths, but same degree of phase synchronization, exhibited different dynamical behaviors (reflected in their promiscuity), illustrating that the synchronization behavior of a network goes beyond just the degree of synchronization. 

We conclude that spatial variability affects not just the degree of phase synchronization, but also its form: it generates promiscuity, influencing the duration of phase-locking. We suggest, then, that spatial variability is a useful quantity to consider in the study of neuronal communication.

\section*{Acknowledgements}
This study was financed in part by the Coordena\c c\~ao de Aper\-feiçoamento de Pessoal de N\'{\i}vel Superior - Brasil (CAPES) - Finance Code 001. The authors also acknowledge the support of  Conselho Nacional de Desenvolvimento Cient\'{\i}fico e Tecnol\'ogico,  CNPq - Brazil, grant number 302785/2017-5, and Finan\-ciadora de Estudos e Projetos (FINEP).

\section*{References}
\bibliographystyle{model1-num-names}
\bibliography{references.bib}

\end{document}